\documentclass[sigconf]{acmart}  \twocolumn

\usepackage{textcomp}
\usepackage{multirow}
\usepackage[ruled,linesnumbered,noend]{algorithm2e}

\usepackage{subcaption}
\usepackage{amsmath,amssymb,amsfonts}
\usepackage{gensymb}
\DeclareMathOperator*{\argmax}{argmax} % thin space, limits underneath in displays
\usepackage[ruled,linesnumbered,noend]{algorithm2e} 
\usepackage{graphicx}
\usepackage{textcomp}

\usepackage{tabularx}
\usepackage{makecell}

\usepackage{mathtools}

\usepackage{url}
\def\BibTeX{{\rm B\kern-.05em{\sc i\kern-.025em b}\kern-.08em
    T\kern-.1667em\lower.7ex\hbox{E}\kern-.125emX}}

\def\acmBooktitle#1{\gdef\@acmBooktitle{#1}}
\acmBooktitle{Proceedings of \acmConference@name
       \ifx\acmConference@name\acmConference@shortname\else
         \ (\acmConference@shortname)\fi}

\copyrightyear{2020} 
\acmYear{2020} 
\setcopyright{rightsretained} 
\acmConference[NANOCOM '20]{The Seventh Annual ACM International Conference on Nanoscale Computing and Communication}{September 23--25, 2020}{Virtual Event, USA}
\acmBooktitle{The Seventh Annual ACM International Conference on Nanoscale Computing and Communication (NANOCOM '20), September 23--25, 2020, Virtual Event, USA}\acmDOI{10.1145/3411295.3411317}
\acmISBN{978-1-4503-8083-6/20/09}

\begin{document}

\title{Ultra-dense Low Data Rate (UDLD) Communication in the THz}
\thanks{This project was funded by CMU Portugal Program: CMU/ECE/0013/2017- THz Communication for Beyond 5G Ultra-fast Networks.}
\author{Rohit Singh}
\email{rohits1@andrew.cmu.edu}
\affiliation{%
  \institution{\textit{Engineering \& Public Policy}, \textit{Carnegie Mellon University}}
  %\streetaddress{}
  \city{Pittsburgh}
  \country{USA}
  %\state{a}
  %\postcode{a}
}

\author{Douglas Sicker}
\email{sicker@colorado.edu}
\affiliation{%
  \institution{\textit{ Computer Science}, \textit{University of Colorado}}
  %\streetaddress{}
  \city{Denver}
  \country{USA}
  }

\maketitle

% The abstract is a short summary of the work to be presented in the article.
\section*{ABSTRACT}

In the future, with the advent of the Internet of Things (IoT), wireless sensors, and multiple 5G  applications yet to be developed, an indoor room might be filled with $1000$s of devices. These devices will have different Quality of Service (QoS) demands and resource constraints, such as mobility, hardware, and efficiency requirements. The THz band has a massive greenfield spectrum and is envisioned to cater to these dense-indoor deployments. However, THz has multiple caveats, such as high absorption rate, limited coverage range, low transmit power, sensitivity to mobility, and frequent outages, making it challenging to deploy. THz might compel networks to be dependent on additional infrastructure, which might not be profitable for network operators and can even result in inefficient resource utilization for devices demanding low to moderate data rates. Using distributed Device-to-Device (D2D) communication in the THz, we can cater to these ultra-dense low data rate type applications in a constrained resource situation. We propose a 2-Layered distributed D2D model, where devices use coordinated multi-agent reinforcement learning (MARL) to maximize efficiency and user coverage for dense-indoor deployment. We explore the choice of features required to train the algorithms and how it impacts the system efficiency. We show that densification and mobility in a network can be used to further the limited coverage range of THz devices, without the need for extra infrastructure or resources.

\section*{Keywords}  Terahertz (THz), Beyond 5G (B5G), Indoor deployment, Densification, Device-to-device (D2D) Communication, Multi-Agent Reinforcement Learning (MARL).

%%%%%%%%%%%%%%%%%%%%%%%%%%%%%%%%%%%%%%%%%%%%%%%%%%%%%%%%%%%%%
%%%%%%%%%%%%%%%%%%%%%%%%%%%%%%%%%%%%%%%%%%%%%%%%%%%%%%%%%%%%%
%%%%%%%%%%%%%%%%%%%%%%%%%%%%%%%%%%%%%%%%%%%%%%%%%%%%%%%%%%%%%

\section{Introduction}

The International Telecommunication Union (ITU) 5G minimum performance requirement report suggests that 5G has to support nearly $1 million$ devices per $km^2$ outdoors and $1000$s of devices per $100m^2$ indoors, with ultra-low latency of nearly $1ms$, reliable communication with low Block Error Rate (BLER) of $10^{-5}$, and low energy consumption \cite{IMT5GReq, TeraCom}. Typical applications would range from the Internet of Things (IoT), ultra high definition streaming, dynamic monitoring, sensor networks, eHealth and body-centric applications, Augmented Reality (AR), Virtual Reality (VR), and other applications yet to be developed \cite{ITU2030, TopDown}. Although 5G is at its nascent stage, it is likely that most of the aforementioned applications and its high demands may not be satisfied by 5G alone. In the future, the network will need to be more tactile, i.e., the network\textquotesingle s response time is ultra-low and can be perceived in real-time. The human-machine interaction and even machine-machine interactions could require response times $<1ms$. Further densification and heterogeneity of User Equipment (UEs) and Access Points (APs) will throttle the network and increase these response times. The network will also need to be efficient (both energy and spectrum) and smart while dividing resources among these dense UEs and APs. This increased complexity has compelled the industry, academia, and regulators to start thinking about Beyond 5G (B5G) applications \cite{OurTPRC, B100GHz}, which is focused on making the network faster, smarter, and more efficient. 

\begin{figure}[t]
\centering
\includegraphics[width=3.5 in,height=2.7 in]{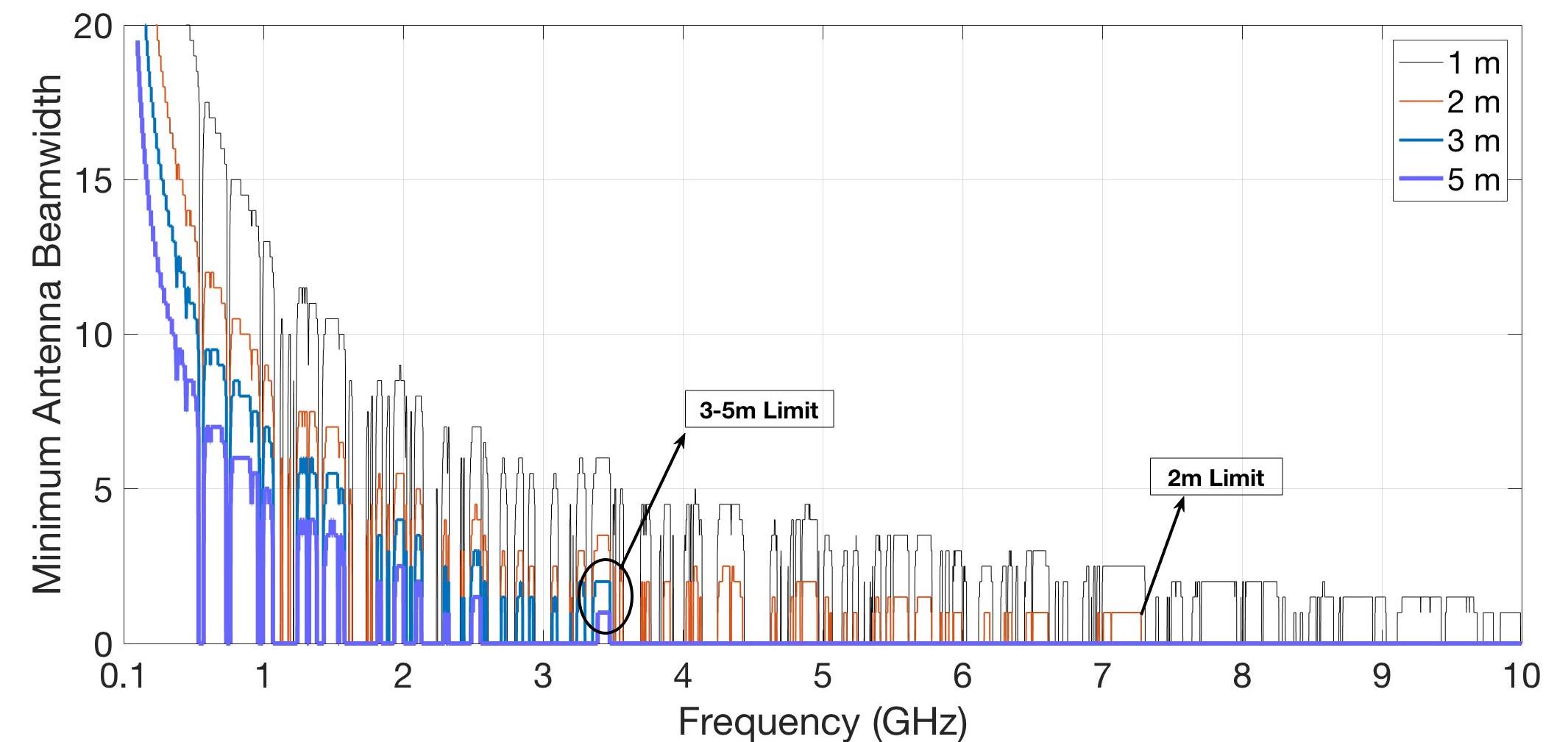}
\caption{Minimum antenna beamwidth (in degree) required in the THz band to achieve a channel capacity of $10$ $bps/Hz$ for static users at $60 \%$ relative humidity and varying distances.}
\label{DistLim}
\end{figure}

One such promising technology is the THz band ($275$ $GHz$-$10$ $THz$) \cite{OurTPRC}, which can provide a massive amount of \textit{greenfield contiguous} spectrum, ultra-high-throughput of $\approx 1$ $Tbps$, reduced harmful interference, and secure type communication. These benefits were not naturally available in the traditional Radio Frequency (RF) and millimeter-Wave (mmWave) bands \cite{TeraCom, TopDown, OurTPRC}. However, THz technology is still at a nascent stage. Given the high spreading, absorption, and penetration losses, communication in the THz spectrum is challenging. To compensate for the losses, devices will need higher antenna gain or narrower antenna beamwidth. Using the antenna gain analysis for static users proposed in \cite{OurGC}, shows that even with the narrowest pencil-beams, there are distance and operating frequency limitations, which makes the THz very sensitive to use-cases, shown in Fig. \ref{DistLim}. Moreover, narrower antenna beamwidth can result in frequent mobility induced outages and increased latency due to the constant need for beam alignment \cite{OurCCNC1, CovTHz}. Thus there is a need for methods that opportunistically communicate in the THz band.

\setlength\belowcaptionskip{-0.2 in}
\setlength{\abovecaptionskip}{0 in}

\subsection{Dense Deployment in THz}

It is envisioned that, in the future, there will be a dense use-case for indoor devices, and a single user or an indoor robot can bear multiple devices or sensors at a time \cite{TeraCom}. For example, a user/robot can have a phone, smartwatch, smart photonic wrist bands, VR glasses, smart clothes, implanted medical devices, and multiple other sensors. This explosion of indoor devices coupled with random small-scale mobility of the user \cite{OurGC} can overwhelm the indoor APs. On the other hand, these devices might not require an ultra-high data rate but need a bare minimum data rate of a few $Mbps$, which can be satisfied by a small slice of the spectrum. Using the lower frequency RF and mmWave bands for this dense deployment is still not feasible due to multiple reasons, such as interference issue, device scalability, need for complex multiple access schemes, need for carrier aggregation, and lack of security. The THz spectrum has enough bandwidth to accommodate the dense deployment of devices. For example, even the smallest frequency windows of $1 GHz$, can effectively accommodate $1000$'s of devices, such as Narrow Band IoT (NB-IoT), Internet of Nano-Things (IoNT), Body Area Network (BAN), Wireless Personal Area Network (WPAN), and Wireless Sensor Network (WSN), by using simple Frequency Division Multiple Access (FDMA) scheme. In this paper we propose the use-case of Ultra-Dense Low Data-rate (UDLD) applications for the THz band.

\begin{figure}[t]
\centering
\includegraphics[width=3.5 in,height=2.5 in]{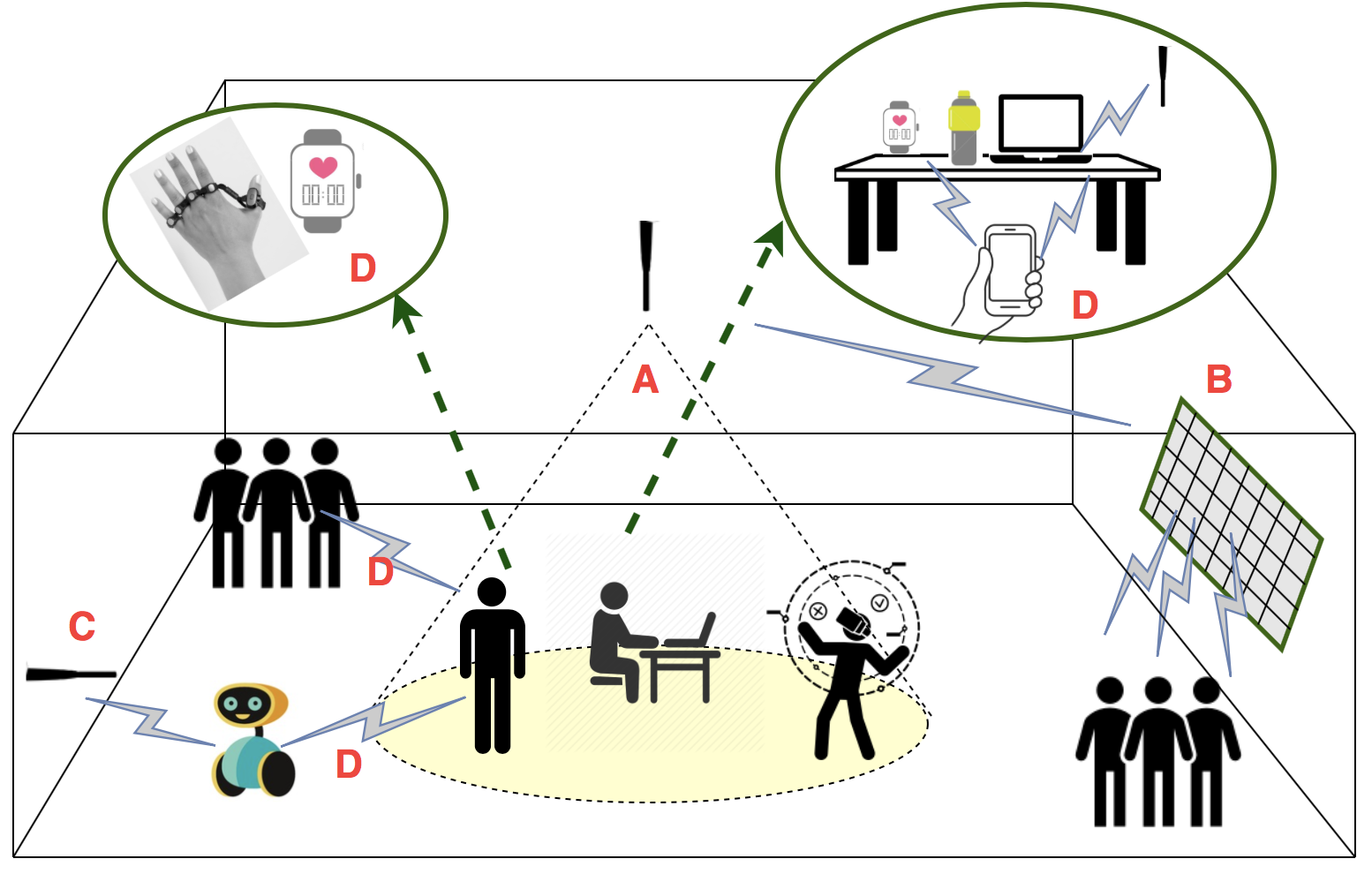}
\caption{Types of communication and use-cases in the THz band, which can be used to improve spatial coverage.}
\label{Moti}
\end{figure}

Densification of devices will not only overwhelm the system, but also can act as blockages for the THz signals. Besides high mobility and frequent change in device orientation, i.e., small-scale mobility \cite{OurGC}, will lead to outages and lower coverage. However, with opportunistic use of the THz band, scheduling, and efficient resource allocation, multiple use-cases requiring a low data rate for indoor devices can be satisfied \cite{SensMOP}. Fig. \ref{Moti} shows four types of communications links for \textit{dense indoor} networks in the higher frequency bands. In type A, the users are at a direct line-of-sight (LoS) with the AP, but the range is limited due to losses and presence of blockages, as emphasized in Fig. \ref{DistLim}. The devices, which are out of the AP's range (shown in yellow), will need additional infrastructure to get coverage. It can be done either through, type B, i.e., intelligent reflecting surface (IRS) \cite{ReconSur,IRS}, or type C, i.e., additional APs deployed closer to the users \cite{ThzPlug, SHINE}. Both types B and C can prove to be costly and inefficient for indoor UDLD use-cases. With densification and increased random user mobility in the system the infrastructure requirement for type B and C will be higher. On the other hand, a fourth method, type D, i.e., Device-to-Device (D2D) communication, can opportunistically cater to this dense deployment. Type D can increase the range of type A communication to reach further in the room by relaying the signals. Moreover, there will be devices which are within the AP's range but are blocked by objects, such as furniture or human body parts, or suffer frequent outages due to small-scale mobility. Using dedicated infrastructure (like, type B and C) for these use-cases might not be efficient, rather type D communication can be used opportunistically to cater to these devices, as shown in  Fig. \ref{Moti}.

\setlength\belowcaptionskip{-0.2 in}
\setlength{\abovecaptionskip}{0 in}

D2D communication has been widely studied in the RF and mmWave bands \cite{D2D2}\cite{D2D3}, and there are multiple challenges associated with using it in the THz band. Due to limited signal coverage, \textit{centralized D2D} communication is challenging. Devices will require a common channel in the lower RF bands to communicate with the AP. Radios having the flexibility to sweep from GHz to THz frequency are hard to manufacture and not a scalable solution for the use-cases mentioned above. A \textit{distributed D2D} approach is a more viable solution. Nevertheless, in distributed D2D for dense mobile networks, finding the optimal path will be complex, and multi-hop communication will be inefficient and unreliable \cite{D2D1}. With proper scheduling \cite{D2D4, D2D5, D2D6}, significant coverage can still be achieved in the THz. In this paper, we propose the use of densification and random user mobility to the network's advantage to improve the coverage range of a THz Access Point (THz-AP). Since UDLD devices do not require a high throughput instead of the traditional allocation method of strongest signal strength, we focus on opportunistic use and smart learning algorithms to predict resource allocation in a distributed D2D. Instead of a multi-hop ad-hoc type communication, we propose a \textit{2-layered} or 2-hop distributed D2D approach for resource allocation and scheduling in the THz and use coordinated multi-agent reinforcement learning (MARL) for the device to efficiently allocate those resources.

%Due to mobility multi-hop D2D will be limited in THz.. In THz …in a dense and mobile network… unnecessary and varying…..But there can be ...Given this form of communication.. it is important to efficiently use the resources.. Putting ... and can result in wastage of resources 

\subsection{Our contribution}
The rest of the paper and the contribution is listed as follows:
\begin{itemize}
    \item We propose a use-case for a dense indoor type deployment in the THz, and show even with densification and random user mobility in the environment the network can still maintain significant coverage for UDLD type applications.
    \item We propose a \textit{2-layered} distributed D2D approach that operates in a constrained resource environment (i.e., a single THz-AP, limited bandwidth, and limited antenna gain). The Layer 1 devices have direct access to the resources, i.e., a direct LOS to the AP. A selected number of devices from Layer 1 act as relays between the Layer 2 device and the THz-AP, explained in Section \ref{LMod}
    \item For efficient resource allocation instead of a central control or transmitting topological information, which will result in increased network load and latency, we allow the Layer 1 device or the agents to work in a coordinated manner. The agents independently explore the environment and make resource allocation decision, and improve their local decisions by coordinating or exchanging feedback from the AP, explained in Section \ref{MarMod}
    \item To enable these agents to make efficient resource allocation they need detailed knowledge about the environment, such as exact location and mobility. However, many of the UDLD devices might not have such sophisticated hardware to obtain location information or can result in significant privacy risks. Thus, we propose two models: D2D Model 1, which uses the Layer 2 device location, and D2D Model 2, which uses number of neighboring agents, as its feature set. The proposed models and evaluation is explained in Sections \ref{d2dMod} and \ref{Eval} respectively. 
\end{itemize}

%%%%%%%%%%%%%%%%%%%%%%%%%%%%%%%%%%%%%%%%%%%%%%%%%%%%%%%%%%%%%
%%%%%%%%%%%%%%%%%%%%%%%%%%%%%%%%%%%%%%%%%%%%%%%%%%%%%%%%%%%%%
%%%%%%%%%%%%%%%%%%%%%%%%%%%%%%%%%%%%%%%%%%%%%%%%%%%%%%%%%%%%%
\section{System Model} \label{Sys}

In this section, we develop the 2-Layer distributed D2D model using coordinated MARL to improve coverage in dense indoor THz. 

\subsection{Layer Model} \label{LMod}
\begin{figure}[t]
\centering
\includegraphics[width=3.5 in,height=2.5 in]{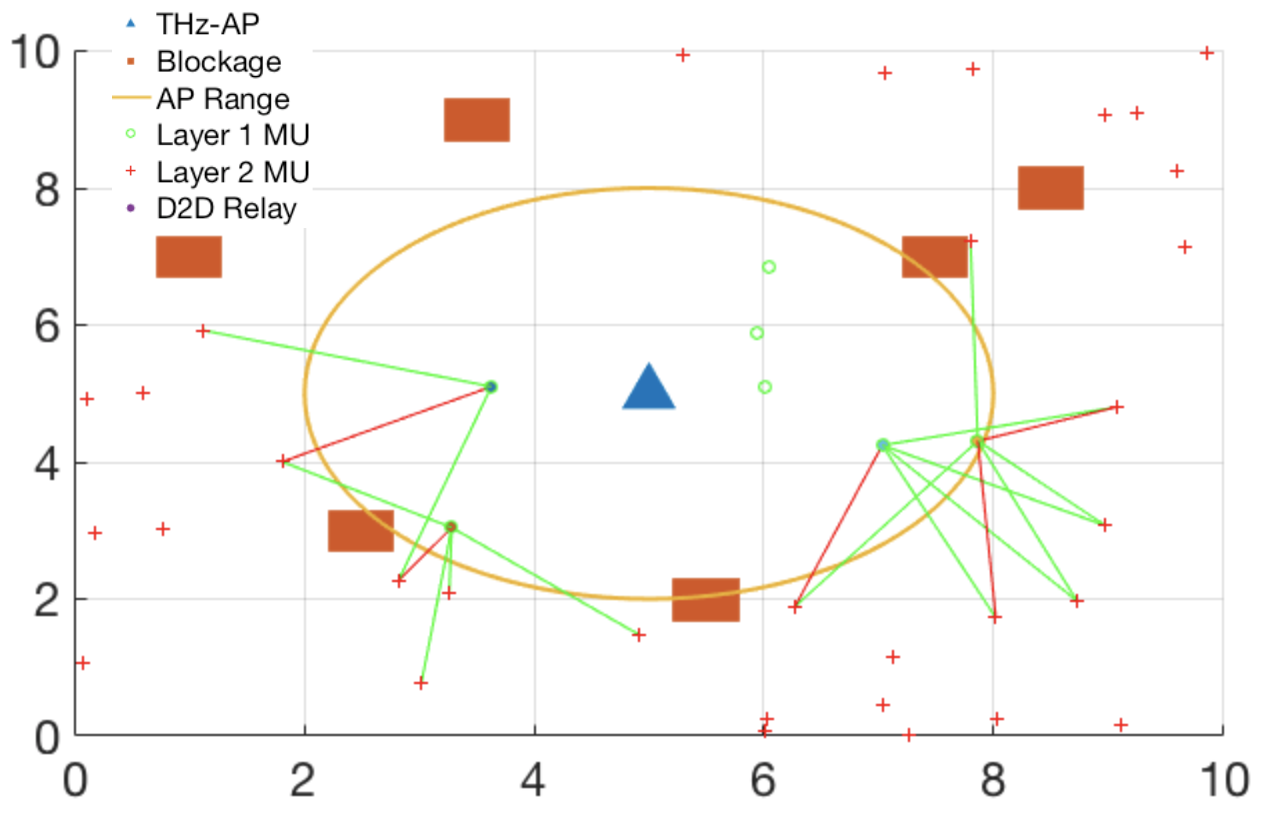}
\caption{A Dense-Indoor Environment with Layer 1 MU acting as relays for the Layer 2 MU. A  Layer 1 MU opting in to serve a particular Layer 2 MU is shown in green and red otherwise.}
\label{Env}
\end{figure}

Let us consider an indoor environment with few static blockages and a single THz-AP located at the center of the room as shown in Fig. \ref{Env}. All the devices are mobile and follow a \textit{random way point} model. Let the set of these Mobile Users (MUs) be denoted as $\mathcal{M}$. The devices that have a direct LoS to the THz-AP are considered as Layer 1 MUs denoted by the set $\mathcal{M}_F$ and the remaining devices are Layer 2 MUs denoted by the set $\mathcal{M}_S$, such that $\mathcal{M}_F \cap \mathcal{M}_S=\emptyset$. The Layer 1 MUs can be considered as part of the type A links shown in Fig. \ref{Moti}. To extend the coverage range of type A communication and reach the devices in $\mathcal{M}_S$, we use D2D communication, which is type D. 

The AP divides the resources equally among the Layer 1 MUs for the devices to independently operate. Thus the bandwidth for the $i^{th}$ device is $B_i=B/|\mathcal{M}_F| \forall i \in \mathcal{M}_F$, where $B$ is the total available spectrum. Moreover, the achievable data rate for these users are dependent on the distance from the AP $d_i$ and can be calculated through Equation (\ref{RateEq}), where $P_t$ is the transmit power, $G_t (\delta)$ and $G_r (\delta)$ is the transmitter and receiver antenna gains respectively, $\delta$ is the antenna beamwidth, $L_A$ is the absorption loss, $L_S$ is the spreading loss or Free Space Path Loss (FSPL), $f_c$ is the operating frequency, and $N_o$ is the noise spectral density in $dB/Hz$. $L_S$ can be calculated based on the frequency and distance, while for $L_A$ we also need to consider the medium absorption coefficient, which is dependent on the relative humidity $\rho$, and room temperature $T$ \cite{OurGC}. Moreover, in a dense deployment interference can cause a significant reduction in signal strength. However, THz offers massive bandwidths, and with proper frequency allocation for UDLD devices, the interference can be ignored.

\setlength\belowcaptionskip{0 in}
\setlength{\abovecaptionskip}{0 in}

\abovedisplayskip=-4pt
\belowdisplayskip=4pt
\begin{eqnarray}
\mathcal{R}_i=B_i log(1+\frac{P_t*G_t (\delta)*G_r (\delta)}{L_A (f_c,d_i,\rho, T)*L_S (f_c,d_i)* N_o*B_i })
\label{RateEq}
\end{eqnarray}

Now using D2D communication, the $i^{th}$ Layer 1 MU can independently reach the $j^{th}$ Layer 2 MU, where $j \in \mathcal{M}_S$. As shown in Fig. \ref{Env} not all Layer 1 MUs are considered as D2D Relays. The Layer 1 MUs, which are at the center of the room are not idle for becoming a D2D relay, due to high path loss. Rather the devices, which are at the edge of the THz-AP range are likely to be candidate for D2D Relay. Furthermore, some $j^{th}$ Layer 2 MUs might have multiple links with some D2D Relay, which might lead to over-coverage and resource wastage. Thus, there exists (a) an optimal set of D2D Relay devices from the Layer 1 MUs, (b) and optimal links between the D2D Relay devices and Layer 2 MUs. The D2D Relay can either serve or not serve a device from Layer 2 MU, which is shown through the red and green links respectively in Fig. \ref{Env}. A combination of the red and the green links that results in maximum coverage and minimum resource allocation can be considered as an optimal link. Due to the dynamic nature of the system the presence of the devices in the two layers will change and so will the optimal links. In a centralized approach the optimal D2D devices and its optimal links can be monitored at the cost of increased network load and resource utilization, which is not the case for distributed systems. For distributed systems, to maximize the system coverage, the choice of these links need to be opportunistic and can be predicted based on the system parameters. Let the set of the optimal links from Layer 1 MUs to Layer 2 MUs be $\mathcal{M}_R \subseteq \{\mathcal{M}_F \text{X} \mathcal{M}_S\}$.

\subsection{MARL Model} \label{MarMod}

Although more red links (shown in Fig. \ref{Env}) will result in lower coverage, while more green links might result in higher coverage, it might also lead to over-coverage or under-coverage at multiple instances. This problem can be solved by exchanging topological information and access control strategies through a central node or multi-hop communication. Due to limited coverage, increased overhead, and increase infrastructure cost, a central D2D or distributed multi-hop D2D is challenging in the THz. Instead we propose a 2-layered distributed D2D as explained earlier. To implement a 2-layered distributed D2D we propose using MARL and allowing the Layer 1 MUs to behave as independent agents and learn the optimal $\mathcal{M}_R$ set for the system. Please note we know refer the Layer 1 MUs as \textit{agents} for the rest of the paper.

To efficiently select the set $\mathcal{M}_R$ for the system we use reinforcement learning. Reinforcement learning (RL) is a machine learning algorithm where an agent learns the optimal policy from interacting with the environment and updating the policy based on the rewards obtained for its state and action pair. This concept can be further extended to a multi agent system where each agent comes up with its own local policy. Instead of fully independent operation of the agents we propose the use of coordinated MARL, where the agents interact with the THz-AP, and share the experience about similar tasks and learn faster \cite{MARL1, MARL2}. Furthermore, due to mobility the sets $\mathcal{M}_F$ and $\mathcal{M}_S$ will be dynamic, and the devices will keep on moving across these sets. Thus, a new agent, will need to learn the system optimal policy from scratch and will require multiple samples. With cooperative MARL, the agents can share their learned policies and help the new agents learn faster. 

\setlength\belowcaptionskip{-0.2 in}
\setlength{\abovecaptionskip}{0 in}
\begin{figure}[t]
\centering
\includegraphics[width=3 in,height=2 in]{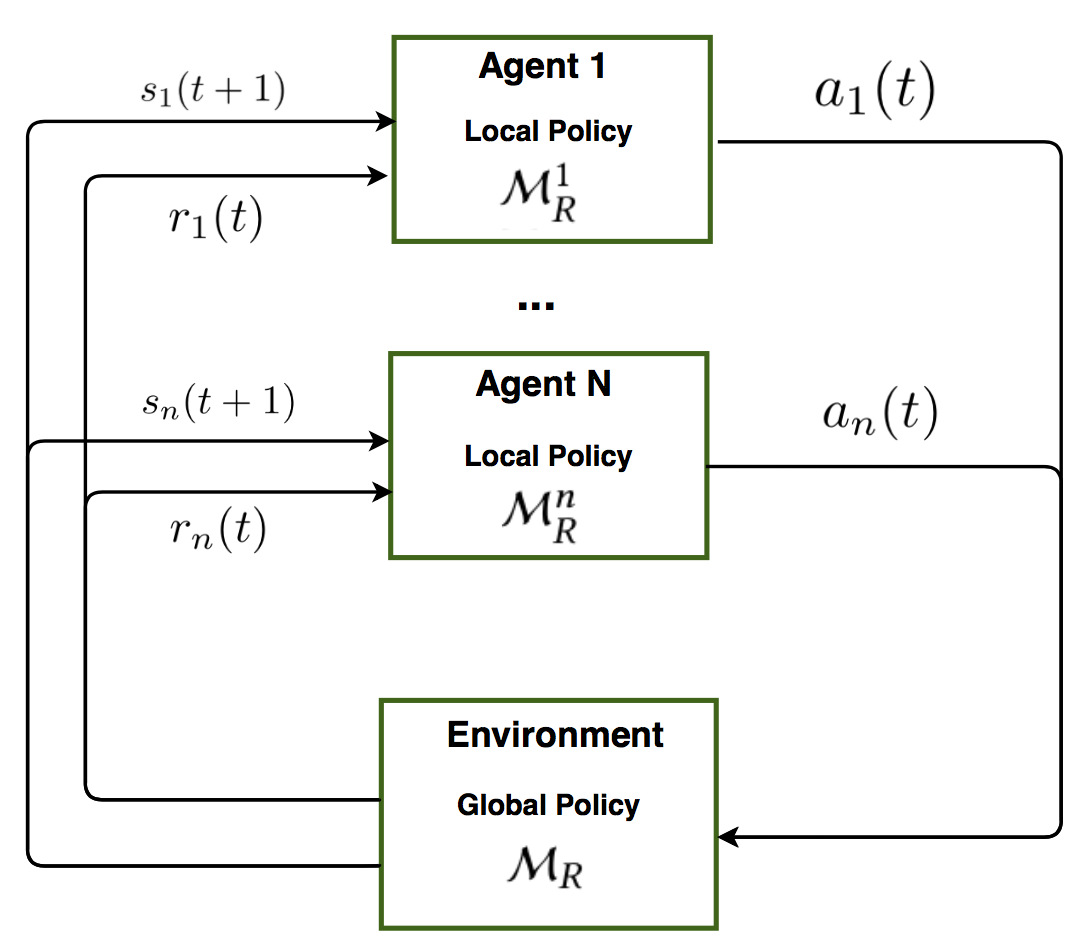}
\caption{MARL model along with the state transitions among agents and the environment.}
\label{MARLMO}
\end{figure}

Now each agent $i$ at a time period $t$ can independently take an action $a_i(t)$ based on the state information $s_i(t)$, where $a \in \mathcal{A}$ and $s \in \mathcal{S}$. The agent then receives the reward $r_i(t)$ from the AP to update its local policy, as shown in Figure \ref{MARLMO}. Most of the cooperative setting involves homogeneous agents with a common reward function. Each agent learns its own policy while making sure that the global policy is maintained, as shown in Fig. \ref{MARLMO} \cite{MARL1, MARL2}. In this problem the optimal local policy for each agent $i \in \mathcal{M}_F$ is $\mathcal{M}_R^i \subseteq \{ \{i\} \text{X} \mathcal{M}_S^i\}$ and the global policy is constructing the set $\mathcal{M}_R=\bigcup\limits_{i=1}^{|\mathcal{M}_F|} \mathcal{M}_R^i$. Thus the action set $\mathcal{A}=\{0,1\}$ consists of two options, either there exists a link from agent $i$ to Layer 2 MU $j$, i.e., $1$ or the link does not exists, i.e., $0$, as shown in Fig.\ref{Moti} with the green and red lines respectively. 

We use Q-learning algorithm for the agents to find the best action $a^*$ given the state $s(t)$. Based on the state action pair we can construct a Q-Table that keeps track of the best Q-value, i.e., the state action pair which returns the best reward. Let us define the Q value for the $i^{th}$ agent as $Q^{(s,a)}_i(t)$. The Q-table is updated at each time step as shown in Equation (\ref{Qup}), where $\alpha$ is the learning rate and $\beta$ is the discount factor:

\abovedisplayskip=1pt
\begin{eqnarray}
\begin{split}
&Q^{(s,a)}_i(t) = Q^{(s,a)}_i(t) + \\
&\qquad\qquad  \alpha \left( R_i(t)+ 
            \beta \max_{a'}Q^{(s,a)}_i(t+1) - Q^{(s,a)}_i(t) \right)
\end{split}
\label{Qup}
\end{eqnarray}

In this problem, the state space $\mathcal{S}$ can incorporate multiple environmental factors, such as location, mobility, neighbor information, and other resources. These factors will result in a huge number of states, and the Q-function can no longer be represented as a table. It will be challenging to implement MARL in such a scenario. Firstly, the sensors and devices will not have so much memory to process such a big table. Secondly, even if the agents have enough memory, the learning process will be slow and require many training data. Therefore, instead of a large table, we parameterize the state space, such that when an agent updates the parameter from its experience, the Q estimate for other parameters will also change. Let us parameterize and redefine the states space $\mathcal{S}$ for agents considering $m$ environmental factors or features as $\phi_{ik}(s)$ where $i \in MU_F$ and $k \in \{1, \cdot \cdot \cdot, m\}$. Let us redefine $Q^{(s,a)}_i(t)$ as  $Q^{(\pmb{\phi(s)},a)}_i(t)$, shown in Equation (\ref{QGen}), where, $\pmb{\phi_i(s)}$ is a vector with all the features of size $1$X$(m+1)$ and $\theta$ is the weight vector that keeps track of the changes in the Q-values for particular parameter and actions. 

\abovedisplayskip=-4pt
\belowdisplayskip=4pt
\begin{eqnarray}
Q^{(\pmb{\phi}_i,a)}_i= \theta_0+\theta_1\phi_{i1}(s)+ \cdot\cdot\cdot +\theta_m\phi_{im}(s) = \pmb{\phi_i(s)^T}\pmb{\theta_i}
\label{QGen}
\end{eqnarray}

The weight vector $\theta$ is of size $(m+1)$X$|\mathcal{A}|$, where the $1$ is for the bias. Therefore in a linear Q-function approximation the agents will be updating the $\pmb{\theta_i}$ for the optimal policy as shown in Equation (\ref{WtUp}).

\abovedisplayskip=-4pt
\belowdisplayskip=4pt
\begin{eqnarray}
\begin{split}
&\theta_i(t) = \theta_i(t) + \\
&\qquad\qquad \alpha \left (R_i(t)+ \beta \max_{a'}Q^{(\pmb{\phi}_i,a)}_i(t+1) - Q^{(\pmb{\phi}_i,a)}_i(t) \right )  \pmb{\phi_i(s)}
\end{split}
\label{WtUp}
\end{eqnarray}

%%%%%%%%%%%%%%%%%%%%%%%%%%%%%%%%%%%%%%%%%%%%%%%%%%%%%%%%%%%%%
%%%%%%%%%%%%%%%%%%%%%%%%%%%%%%%%%%%%%%%%%%%%%%%%%%%%%%%%%%%%%
%%%%%%%%%%%%%%%%%%%%%%%%%%%%%%%%%%%%%%%%%%%%%%%%%%%%%%%%%%%%%
\section{D2D-Relay Node Selection Models} \label{d2dMod}

In this section, we discuss the agents learning procedure and its interaction with the environment and the THz-AP for coordination. 

\subsection{Agent Learning}

\begin{figure}[t]
\centering
\includegraphics[width=3.5 in,height=1.8 in]{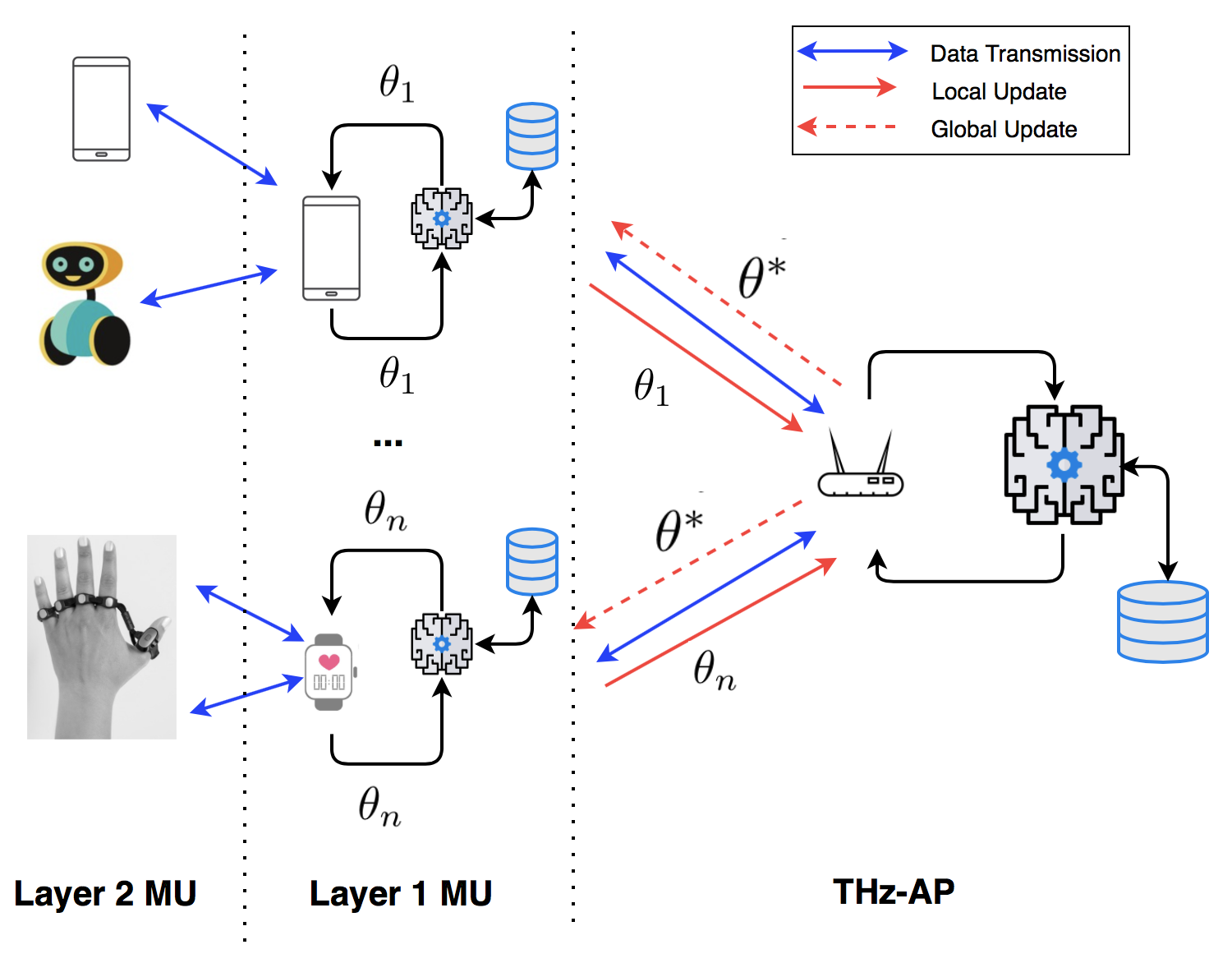}
\caption{Data and policy transmission across layers.}
\label{LayCom}
\end{figure}

The agents act as a bridge between the THz-AP and the Layer 2 devices, as shown in Fig. \ref{LayCom}. The agents transmit data and learning parameters across the layers. Moreover, these agents also have the cognitive capacity to make smart decisions, which the relatively simple Layer 2 sensors cannot perform. 

For the agents to efficiently update its $\pmb{\theta}$ vector, it needs proper feedback from the system or rewards. In this problem, we use a private and public reward scheme for the agents to make optimal decisions. The private reward is a function of the throughput $\mathcal{R}_{ij}$ between the agent $i$ and device $j$, which is based on the action. For an action $a_i=0$ the agent gets $0$ private reward. Otherwise, the reward is calculated using the Equation (\ref{RateEq}). The agents must learn how their individual actions impact the actions taken by other agents. Thus the agents are also provided with a public reward $\mathcal{R}_0$, which is the system throughput, and it changes based on the collective actions considered by the agents. Both the rewards are normalized to a maximum achievable data rate of $\mathcal{R*}$. An agent is provided a positive reward for its optimal action and negative rewards otherwise, which is tracked by the AP. Let $c_j$ be a Boolean variable, which keeps track of Layer 2 coverage, where $1$ means it is already covered by some agent $i$. The AP updates this variable $c_j$ based on the actions it receives back from the agents. A combination of $a_i$ and $c_j$ can result in four types of rewards, to balance under-coverage and over-coverage among agents, as shown in Equation (\ref{RewE}). For example, if an agent decides not to cover a Layer 2 device and the device has already been covered by some other agent, it will get a positive public reward. However, if that device was not covered the agent will get a negative public reward for under-coverage. Similarly, explanation for the agent, which considers $a_i=1$.

\abovedisplayskip=-4pt
\belowdisplayskip=4pt
\begin{eqnarray}
\centering
R=\begin{cases} (\mathcal{R}_{ij} + \mathcal{R}_0)/\mathcal{R^*}  &  
                                                a_i=1 \textrm{\&} c_j=0\\ 
                 (-\mathcal{R}_{ij} - \mathcal{R}_0)/\mathcal{R^*}  &  
                                                a_i=1 \textrm{\&} c_j=1\\ 
                \mathcal{R}_0/\mathcal{R^*} &  
                                                a_i=0 \textrm{\&} c_j=0\\ 
                -\mathcal{R}_0/\mathcal{R^*}  & \textrm{else} 
    \end{cases}
\label{RewE}
\end{eqnarray}

\setlength\belowcaptionskip{0 in}
\setlength{\abovecaptionskip}{0 in}

Pseudocode for an agent learning from the environment is shown in Algorithm \ref{Algo1}. A newly selected agent initializes generalized the Q function as $0$, which builds up either through environment interaction or through policy update with the AP. At Line \ref{SigChk}, the device checks for a minimum signal strength $\gamma_0$ from the THz-AP. If there is no signal, then the device is a Layer 2 MU and waits for an agent to get coverage. Otherwise, the device can act as an agent. The agent iterates for devices in its proximity, i.e., $\mathcal{M}_S^i$, and selects an action for each device based on $\epsilon$-\textit{greedy-decay}. In Line \ref{EnvRead}, the agent interacts with the environment and AP to obtain the next sate information and the rewards respectively and keeps on updating its local policy $\pmb{\theta_i}$.

In Line \ref{GlbUp}, the agents exchange their $\pmb{\theta_i}$ values so that other agents can learn an optimal policy and the new agents can learn faster based on  experience gathered from other agents \cite{MARL3, MARL4}, as shown in Fig. \ref{LayCom}. There are multiple ways to find the optimal policy among agents, for simplicity we consider the policy which gives the maximum reward, i.e., $\theta^*(a)= \theta_p(a), \forall a \in \mathcal{A}, p=argmax_a \left (R_1, \cdot \cdot \cdot, R_n \right )$.

\setlength{\textfloatsep}{0in}% Remove \textfloatsep
\setlength{\intextsep}{0pt}% Remove \textfloatsep
\begin{algorithm}[t]
{\small
  \Begin
  {
 	Set $Q^{(\phi_{ik}(s),a)}_i=0, \forall k \in \{1, \cdot \cdot \cdot, m\}, a \in \mathcal{A}$ \\
	
	\For {$t< \text{Total Time}$}
	{ 
    	\If {SignalReceived($\gamma_0$) \label{SigChk}} 
    	{
		\For {$j \in \mathcal{M}_S^i$}
		{
		$ a_{ij}(t)=\begin{cases} Random(a \in \mathcal{A}) & 1-\epsilon \\
					\argmax_a  Q^{(\pmb{\phi},a)} & \epsilon
             			\end{cases} $ \\
		
		$\pmb{\phi_i}(t+1), r_{ij}(t)=$ Environment($a_{ij}(t)$) \label{EnvRead}\\
		
		Update $\pmb{\theta_i}$ \text{from Equation (\ref{WtUp})} \\
		
		$R_i(t)=R_i(t)+ r_{ij}(t)$ \\
		}
		
		$\pmb{\theta_i}$ =PolicyUpdate($\pmb{\theta_i}$) \label{GlbUp}\\
	    }
	    \Else
    	{
    	WaitForAgent()
    	}
	}
} 
\caption{Learning procedure at the  $i^{th}$ Agent}
\label{Algo1}
}
\end{algorithm} 

\subsection{Feature Selection}

In Section \ref{MarMod}, we introduced that the agents' states can be based on multiple environmental factors, such as location, mobility, neighbor information, and other resources. Many of these factors are challenging to achieve or depend on in a dynamic setting. For example, user location and mobility data can be effectively used to make optimal resource allocation. However, localization in higher frequency bands is still a challenge \cite{IndLoc, IndLoc2}. Estimating distance devices will require either (a) accurate synchronization of the clocks between receivers and transmitters, (b) additional sophisticated hardware to send signals at different velocities, or (c) support from other APs to triangulate the location. Moreover, sharing information about other agents can lead to security and privacy concerns. Privacy is already a challenge due to the densification and ubiquity of IoT and sensors. Moreover, most future sensor devices might use eHealth and body-centric devices, which can cause privacy harm. For example, location information can be used to jam signals in the THz \cite{THzBlk}. Nevertheless, an abstraction or anonymized version of this sensitive data can be shared among the agents through the AP.  For example, instead of the exact distance, devices can share information like the number of devices in their vicinity. This kind of information, though abstract, can help the agent learn faster. 

In D2D Model 1, we propose two factors, (a) distance from the agent to the device $j$ in Layer 2, and (b) the coverage capacity of a device or the queue length. It might be challenging to expect such sophistication from devices, such as simple wrist bands, NB-IoT, or even body-centric devices. So we consider a second model, i.e., D2D Model 2, which uses (a) the number of neighboring agents within a distance $d_0$, which the AP can provide, and (b) the device queue length. We assume the maximum queue length for a device to be $5$; however, it can be based on the hardware specifications, device capacity, or can be instantaneous based on the resource allocated.

%%%%%%%%%%%%%%%%%%%%%%%%%%%%%%%%%%%%%%%%%%%%%%%%%%%%%%%%%%%%%
%%%%%%%%%%%%%%%%%%%%%%%%%%%%%%%%%%%%%%%%%%%%%%%%%%%%%%%%%%%%%
%%%%%%%%%%%%%%%%%%%%%%%%%%%%%%%%%%%%%%%%%%%%%%%%%%%%%%%%%%%%%
\section{Evaluation} \label{Eval}

In this section, we evaluate our proposed D2D models for multiple user distribution. 

\subsection{Environment}

We consider a constraint resource environment with a single AP along with static blockages in a room $10m$X$10m$ in dimension, as shown in Fig. \ref{Env}. We assume the devices to be humans that are walking fast and move in a random way point model \cite{OurGC}. We also consider these devices as blockages and assume an average human shoulder width of $40cm$ as the width of the mobile blockages. All the devices can transmit at a maximum power of $0$ $dBm$ and have access to and limited antenna gain of $27$ $dBi$ or antenna beamwidth of $10 \degree$. We assume all the devices in the room can operate in a fixed frequency range $570$ $GHz$ to $580$ $GHz$ and can tune to smaller channels based on the frequency allocation, which is a based on the number of agents. Based on the antenna beamwidth and operating frequency all the THz devices in this simulation can have a maximum range of $d_0=3m$, as shown in Fig.\ref{DistLim}. Thus, for a room size of $10m$X$10m$ a 2-hop D2D communication should be enough to reach all the devices in the room. However, the presence of static and mobile blockages requires the need for more hops, which can be mitigated through proper resource allocation and smart learning models. We define the total time for all the agents to establish links based on the layer model, transmit data, and update their respective policies as an episode. For the proposed learning models we assume $\mathcal{R*}=10 Gbps$,$\alpha=0.01$, $\beta=0.7$, and $\epsilon=0.5$. For an $\epsilon$-\textit{greedy-decay} we assume that there is $10\%$ decay for every $50^{th}$ episode.

\setlength\belowcaptionskip{0 in}
\setlength{\abovecaptionskip}{0 in}
\begin{figure}[t]
\centering
\includegraphics[width=3.5 in,height=1.8 in]{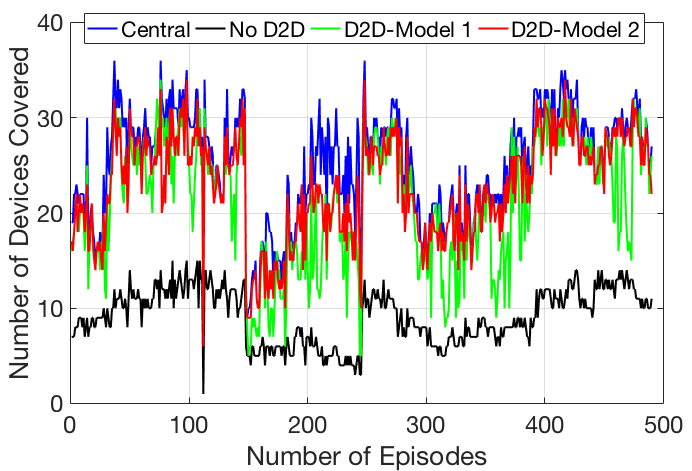}
\caption{Total system coverage with time.}
\label{Cov}
\end{figure}

\subsection{Model Comparison}

\begin{figure}[t]
\centering
\includegraphics[width=3.5 in,height=1.8 in]{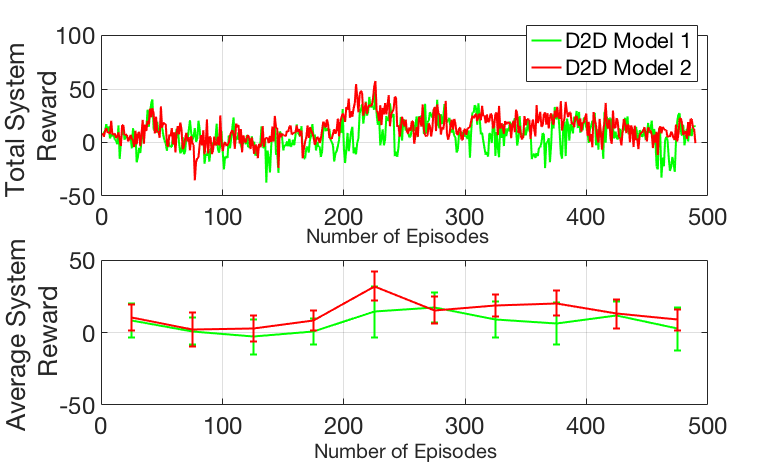}
\caption{Total and average system reward with time.}
\label{Rew}
\end{figure}

We compare our proposed models with two base cases: (a) Central, where the devices share the topological information with the THz-AP and perfect link allocation is maintained, and (b) No D2D, where the users can only be covered if it has a direct LoS with the THz-AP. The Central approach is the best case, and No D2D is the worst case in our setting. As shown in Fig. \ref{Cov}, for $N=40$, the D2D Model 1 and Model 2 are almost close to the Central approach for all the episodes. The variations in the coverage value are due to the dynamic nature of the devices. Model 2 has slightly higher coverage compared to Model 1, which is due to the higher rewards, as shown in Fig. \ref{Rew}. For D2D Model 2, the deviation in rewards is lower compared to the D2D Model 1. This is due to the choice of the feature set. Obtaining information about the number of neighboring nodes helps agents avoid unnecessary negative rewards. As shown in Fig. \ref{Env}, multiple agents try to cover the same Layer 2 device, which results in a negative public and private reward. 

Although Model 2 shows higher coverage compared to Model 1, in real-life, one can be indifferent while choosing between the models. The feature used in Model 1 demands more infrastructure, while the feature in Model 2 increases the risk of privacy violations. The choice of the model can be contextual and is dependent on the availability of hardware and the quality of service demand. Moreover, other features, such as battery power, mobility type, and varying antenna gain, can be used to maximize other system objectives, such as energy, handoff, and latency.

\setlength\belowcaptionskip{0 in}
\setlength{\abovecaptionskip}{0 in}

\subsection{Summary}

\begin{figure}[t]
\centering
\includegraphics[width=3.5 in,height=1.7 in]{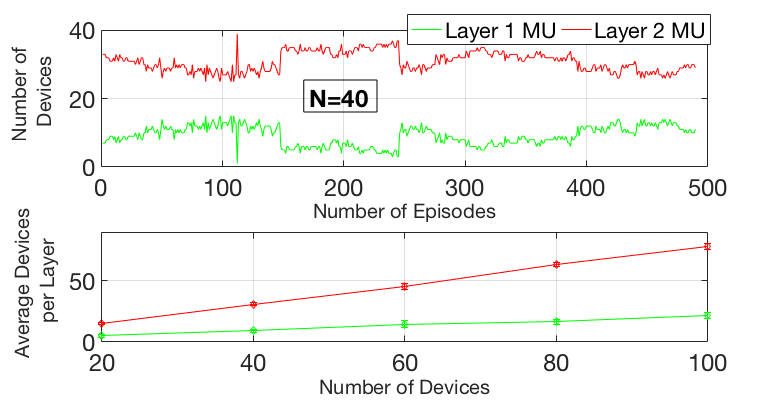}
\caption{Number of Layer 1 and Layer 2 devices based on densification.}
\label{Lay}
\end{figure}

\begin{figure}[t]
\centering
\includegraphics[width=3.5 in,height=1.7 in]{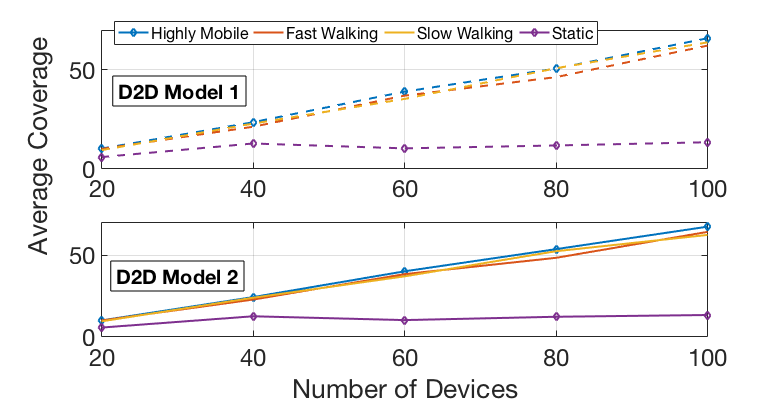}
\caption{Average System Coverage for varying device density and mobility.}
\label{MobDen}
\end{figure}

Since the system is dense and mobile, the coverage and rewards are opportunistic, and dependent on the cardinality of the sets $\mathcal{M}_F$ and $\mathcal{M}_S$. The number of agents are limited for a less dense and relatively static user set. Moreover, in a dense scenario, the devices will act as blockages, which again results in fewer agents, as shown in Fig. \ref{Lay}. If we compare Fig. \ref{Lay} and \ref{MobDen}, despite the lower number of agents, both the proposed D2D models can maintain higher system coverage. However, when densification is coupled with higher mobility, effective system coverage increases. It is only in the case of static users where the coverage is the least because, with higher mobility, the dynamics of the system and the link adjacencies change, creating multiple opportunities throughout the episode where the agents can perform D2D relaying even in a limited coverage scenario. 

The model presented in this paper is based on a 2-Layered approach, which can be extended to a \textit{N-Layered} model. However, it can be argued that increasing the layers might result in higher latency reduced coverage. Our approach constrained the cognitive load to Layer 1, but for an N-Layered model, all of the devices are required to have the processing capacity, which might be costly and inefficient. In our proposed MARL model, the optimal global policy for a particular action $a$ is updated based on the agent with the best reward for that action. Since the optimal link assignments are a function of other spatial parameters, a loose global policy update can lead to over-coverage and under-coverage for different agents. One solution is for agents to have weights between the local and the global policy, where the weight can be learned using simple neural networks.

\setlength\belowcaptionskip{-0.2 in}
\setlength{\abovecaptionskip}{0 in}

\setlength\belowcaptionskip{0.1 in}
\setlength{\abovecaptionskip}{0 in}

%%%%%%%%%%%%%%%%%%%%%%%%%%%%%%%%%%%%%%%%%%%%%%%%%%%%%%%%%%%%%
%%%%%%%%%%%%%%%%%%%%%%%%%%%%%%%%%%%%%%%%%%%%%%%%%%%%%%%%%%%%%
%%%%%%%%%%%%%%%%%%%%%%%%%%%%%%%%%%%%%%%%%%%%%%%%%%%%%%%%%%%%%

\section{Conclusion} \label{Con}

In this paper, we go beyond the traditional view of using THz for ultra-high-throughput to explore dense indoor deployment. We propose a 2-Layered distributed D2D communication for the UDLD type applications in the THz. We use MARL to assign connections between Layer 1 and Layer 2 devices efficiently. We also show that the choice of features for the learning algorithm is subject to hardware availability, privacy risks, and the quality demand. Although mobility has its own set of challenges in THz, mobility when coupled with densification, can help improve coverage in D2D-THz. In a static environment, when the devices are not so mobile, there are fewer chances of efficient D2D communication. However, with more mobility, there will be multiple opportunities throughout the time instance when D2D relaying is possible in a dense environment.

\end{document}